\documentclass[11pt,twoside]{article}

\usepackage{amsmath, amsfonts, amssymb}

\usepackage{graphics}

\begin{document}
\title{PECULIAR FEATURES OF THE VELOCITY FIELD OF OB ASSOCIATIONS
\\ \protect AND THE SPIRAL STRUCTURE OF THE GALAXY}
\author{A.M.Mel'nik}
\date{ \small \it Sternberg Astronomical Institute, Moscow,  Russia
\\ anna@sai.msu.ru
\\ Astronomy Letters,  2003, Vol. 29, pp. 304-310  }
 \maketitle

\begin {abstract}
Some of the peculiar features of the periodic velocity-field
structure for  OB associations can be explained by using the
Roberts-Hausman model, in which the behavior of a system of dense
clouds is considered  in a perturbed potential. The absence of
statistically significant variations in the azimuthal velocity
across the Carina arm, probably, results from  its sharp increase
behind the shock front, which is easily blurred by distance
errors. The existence of a shock wave in the spiral arms and, at
the same time, the virtually free motion of OB associations in
epicycles can be reconciled in the model of particle clouds with
a mean free path of $0.2\text{--}2$~kpc. The velocity field of OB
associations exhibits two appreciable nonrandom  deviations from
an ideal spiral pattern: a 0.5-kpc displacement  of the Cygnus-
and Carina-arm fragments from one another and a weakening  of the
Perseus arm in  quadrant III. However, the identified fragments
of the Carina, Cygnus, and
Perseus arms do not belong to any of the known  types of spurs.
\linebreak spiral pattern, OB associations.

\end {abstract}
\newpage

\section*{Introduction}

Here, we analyze some of the peculiar features or defects of the
periodic  velocity-field structure of OB associations (Mel'nik et
al.~2001). First, we found no statistically significant
variations in the azimuthal  residual velocity across the Carina
arm, although clearly see its variations across the Cygnus and
Perseus arms. Second, OB associations exhibit the features of
both collisional and collisionless motion. Third, the
displacement of individual arm fragments from one another such
that they do not form a single smooth  spiral arm and the gaps in
the spiral pattern should be explained.

Some of the peculiar features listed above can be explained  by
using the model of Roberts and Hausman~(1984), in which the
behavior of a system of dense clouds is considered  in a perturbed
potential. There is ample theoretical and observational evidence
for the existence of small, unaccounted molecular gas
condensations, from which giant molecular complexes are formed at
certain times (Solomon et~al.~1985; Pringle et~al.~2001).

The response of a subsystem of elementary particle clouds to the
propagation of a density wave over  the stellar disk was first
considered by Levinson and Roberts~(1981). They assumed that
clouds could collide inelastically with each other and coalesce
in some cases. In contrast to a continuous gaseous medium, which
responds to a potential perturbation   by an abrupt
 jump of the density and velocity, a subsystem of dense particle clouds exhibits
smoother  variations in gasdynamical parameters across spiral
 arms (Levinson and Roberts~1981).

Roberts and Hausman~(1984) considered the behavior of a subsystem
of clouds with a velocity dispersion of $\Delta u=6$~km s$^{-1}$
and showed that the local frequency of cloud-cloud collisions
increases sharply in spiral arms due to the crowding of cloud
orbits in arms. This increase inevitably gives rise to a shock
wave even in systems with a  large cloud mean free path (see case
$G$, $p=2$~kpc in the above paper).

The study of molecular clouds in the Galaxy is complicated  by
the fact that, in general, we do not know their distances  and
observe their distribution  in the $(l, V_{\textrm{LSR}})$ plane.
However, the spiral structure of the Galaxy is  difficult to
analyze in the $(l, V_{\textrm{LSR}})$ plane and  virtually
impossible in the inner regions (Kwan and Veides~1987; Adler and
Roberts~1992). Constructing and analyzing  the velocity field of
young stars, which, on  average, have the same velocities as
their parent molecular clouds, give a unique opportunity for
studying the velocity field of molecular clouds and the Galactic
spiral structure.

\section*{Analysis of the Velocity Field of OB Associations}

\subsection*{The Periodic  Velocity-Field Structure of OB Associations}

Analysis of the velocity field of OB associations within 3~kpc of
the Sun revealed periodic variations in the radial component
$V_R$ of residual velocity along the Galactic radius vector with
a  scale length of ${\lambda=2\pm0.2}$~kpc and amplitude
$f_R=7\pm2$~km s$^{-1}$ (Mel'nik et al.~2001). Figure~1a shows
the distribution of radial  residual velocities $V_R$   of OB
associations along Galactocentric distance. The filled and open
circles represent the associations located in the regions
${0<l<180^\circ}$ and ${180<l<360^\circ}$, respectively. This was
done in order to separate the initial parts of the arms in
quadrants I and II from their extensions in  quadrants III and
IV. The periodic variations of the two  velocity components were
assumed to be in the shape of sine waves. The parameters of the
sine waves were determined  by solving the equations for the
line-of-sight velocities and proper-motions of OB associations in
two regions: ${30<l<180^\circ}$ and ${180<l<360^\circ}$ (for more
details, see Mel'nik et al.~2001 ). The minima in the radial
residual velocity distribution determine the kinematical
locations of the arm fragments, which we call the Carina (open
circles, $R=6.5$~kpc), Cygnus (filled circles, $R=6.8$~kpc), and
Perseus (filled circles, $R=8.2$~kpc) arms.

In Fig.~1a we see no second minimum in the radial  velocity
distribution  in the region  ${180<l<360^\circ}$ (open circles).
This second minimum should have corresponded to the extension of
the Perseus arm in  quadrant  III. The local decrease in $V_R$ is
observed at a Galactocentric distance of $R=8.5$~kpc, but it does
not reach negative values.

As regards the azimuthal velocity field, periodic variations in
the azimuthal component $V_\theta$ of residual velocity  with an
amplitude of $f_\theta=7\pm2$~km s$^{-1}$ are observed in the
region ${30<l<180^\circ}$ where the Cygnus and Perseus arms are
located (Fig.~1b).  The positions of the minima in the
distributions  of the radial (Fig.~1a) and azimuthal (Fig.~1b)
residual velocities, which determine the kinematic locations of
the Cygnus and Perseus arms, coincide,  within the error limits.
At the same time, no significant variations in the azimuthal
residual velocity were found   in the region ${180<l<360^\circ}$,
where the Carina arm lies and where the imaginary extension of
the Perseus arm could be located.

The interarm distance of $\lambda=2$~kpc yields a mean arm pitch
angle of $i=5^\circ$ for the two-armed model of  the spiral
pattern. The Galactocentric distance of the Sun was assumed to be
 ${R_0=7.1}$~kpc (Rastorguev et al.~1994; Dambis et al.~1995;
Glushkova et al.~1998).

\subsection*{The Profiles of  Cloud-Velocities Variations\protect\\
in a Shock Wave}

The coincidence of the positions of the minima in the  radial and
azimuthal velocity distributions in the Cygnus and Perseus arms
suggests the existence  of a shock wave and cloud-cloud
collisions. In a shock wave, extremely negative values of the
perturbations of the two  velocity components must be reached at
the same phase at the shock front (Roberts~1969; Roberts and
Hausman~1984), while for a  collisionless system the minima in
the radial and azimuthal velocity distributions must be shifted by
 $\pi/2$ in wave phase  (Lin et al.~1969). For $\lambda=2$~kpc a phase
change of $\Delta\chi=\pi/2$ corresponds to a 0.5-kpc
displacement across the arm and can undoubtedly be detectable.

Consider  the variations in the two components of the mean cloud
velocity perpendicular ($U_\perp$) and parallel ($U_\parallel$)
to the shock front with wave phase as obtained by Roberts and
Hausman~(1984) for models of cloud subsystems with different mean
free paths: $p=0.2$~kpc and $p=1.0$~kpc (Fig.~2ab). For tightly
wound arms, the Galactic radius vector is almost perpendicular to
 the spiral arms. Therefore, the variation in the mean velocity of a
particle ensemble perpendicular to the shock front (${U_\perp}$),
which is commonly used in gas dynamics, is identical to the
variation in the radial  residual velocity, implying that
$V_R=U_\perp + const$. A similar relation, $V_\theta=U_\parallel
+ const$, also holds  for the velocity component $U_\parallel$
parallel to the shock front whose  variation corresponds to the
variation in  the azimuthal  residual velocity. The wave phase
changes in the direction perpendicular to the spiral arm and for
tightly wound arms this direction virtually coincides with the
direction of the Galactic radius vector. A change in the wave
phase $\chi$ from $0^\circ$ to $360^\circ$ corresponds to an
increase in Galactocentric distance $R$ by an amount close to
$\lambda$.

As we see  from a comparison of Figs.~2a and 2b, the profiles of
the  cloud-velocities variations become more symmetric and the
positions of the  minima in the radial and azimuthal velocity
distributions are shifted from each other with  increasing mean
free path (i.e., with decreasing importance of collisions).
Cloud-cloud collisions cause  the asymmetric changes in the
profiles of the radial and azimuthal velocities, leading to the
coincidence of their minima at  the shock front (Roberts and
Hausman~1984).

Behind the shock front, where star formation is most intense, the
azimuthal velocity varies with  Galactocentric distance  twice as
fast as does the radial velocity (Fig.~2a). Indeed, at the shock
front (${\chi=180^\circ}$) the two velocity components reach
their minima and at the conditional outer arm boundary
(${\chi=270^\circ}$), the azimuthal component already reaches its
maximum, whereas the radial component increases only to  zero
(the mean  between the peaks in Fig.~2a). Since the abrupt
velocity variations   with distance are easily blurred  by
distance errors,  the variations in the azimuthal component
behind the shock front   are more difficult to detect. This may
be the reason  why we found no statistically significant
variations in the azimuthal  residual velocity across the Carina
arm, although the variations in the radial component across this
arm are clearly seen. The variations in the azimuthal velocity
across the Carina arm can be simply blurred by distance errors.
This removes  the serious logical contradiction, because the
absence of variations in azimuthal velocity  across the Carina
arm cannot be explained by any theoretical models.

The fact that we nevertheless see the variations in the azimuthal
velocity  across the Cygnus and Perseus arms can be explained by
their privileged positions. The Perseus arm is located in the
outer region of the Galaxy where objects with erroneous distances
have strongly different line-of-sight velocities and can be
eliminated from the sample. This cannot be done for objects of
the Carina arm ($l=280\text{--}310^\circ$), because the
line-of-sight velocity variations  with distance in this
direction are small. The same difficulty also concerns the Cygnus
arm ($l=75\text{--}100^\circ$), but it is, on  average,  closer
to the Sun than the Perseus and Carina arms, which significantly
facilitates its study.

\subsection*{The location  of the Corotation Radius}

The fact that radial  residual velocities of most ($70\%$) of the
rich OB associations are directed toward the Galactic center
definitely indicates that the region under study is located
within the corotation radius (Mel'nik et al.~2001). We consider
OB associations with more than 30 stars  in the catalog of OB
associations by Blaha and Humphreys~(1989) to be rich. The model
of particle clouds allows us to easily explain the logical
relationship between the kinematics of rich OB associations and
their location relative to the corotation radius. Let us assume
that the star formation in rich OB associations was more intense
precisely because of the increase in the frequency  of
cloud-cloud collisions. The locations of rich OB associations
must then coincide with the location of the shock front. And only
within the corotation radius the location of the shock front
corresponds to the minimum in the cloud radial velocity
distribution. Therefore, all of the  region under study must be
located within the corotation radius or, in other words, the
corotation radius in the Galaxy must lie beyond the region under
study, i.e., beyond the Perseus arm. This implies that the
angular velocity of the spiral pattern  $\Omega_p$ must be lower
than the mean angular velocity of the Galactic rotation at the
Perseus-arm distance. Taking the rotation-curve parameters from
Mel'nik et al.~(2001), we obtain the following constraint on
$\Omega_p$ in absolute units: ${\Omega_p<25}$~km s$^{-1}$
kpc$^{-1}$.

\subsection*{The Mean Free Path of Molecular Clouds and the Axial Ratio of the
Velocity Ellipsoid for  OB Associations}

An analysis of the kinematics of OB associations (Dambis et
al.~2001) showes that the axial ratio of the velocity ellipsoid
for OB associations in the radial and azimuthal directions  is
${\sigma_u:\sigma_v=8.2:5.8=1.4}$. This value is close to the
ratio of the amplitudes of the radial and azimuthal velocities
for epicyclic motions, $2\Omega/\kappa=1.6$, where $\kappa$ is
the epicyclic frequency. An axial ratio of the velocity ellipsoid
close to the Lindblad ratio $2\Omega/\kappa$ was also obtained
for  young clusters  (Zabolotskikh et al.~2002), implying that the
clouds out of which young stars are born move almost freely in
epicycles. Therefore, the clouds collide only at the shock front,
whereas behind the front and in the interarm space they undergo no
collisions and move ballistically. This implies that the cloud
mean free path  behind the shock front must be larger than the
epicycle scale size. For a subsystem with a velocity dispersion
of $\Delta u=6$~km s$^{-1}$ the epicycle radius  in an
unperturbed region of the Galactic disk is  $X\approx0.2$~kpc.
Therefore, for a mean free path $p>0.2$~kpc, cloud-cloud
collisions will occur mostly at the shock front, which is also
confirmed by the direct computations of cloud orbits performed by
Hausman and Roberts~(1984). Note that local value of a free path
depends on cloud density, and its values in the arm and the
interarm region can differ by more than a factor of 6 (Levinson
and Roberts~1981).

\subsection*{Deviations from an Ideal Spiral Pattern}

Figure~3 shows  the velocity field of OB associations in
projection onto the Galactic plane. The circular arcs in this
figure correspond to the minima in the radial (solid line) and
azimuthal (dashed line) residual velocity distributions  and
determine the kinematic positions of the Carina, Cygnus, and
Perseus arms. Both  deviations from an ideal spiral pattern
--- the displacement  of the Cygnus and Carina arm fragments relative to each
other such  that they do not form a single, smooth spiral arm and
the absence of the Perseus arm extension in   quadrant III
--- are clearly seen in Fig.~3. The extension of the Perseus
arm in  quadrant  III shows up neither in kinematics nor in an
increase in the density of young stars.

The kinematic position of the Carina arm \linebreak ($R=6.5\pm
0.1$~kpc) is displaced relative to the Cygnus arm (${R=6.8\pm
0.1}$~kpc) by 0.3~kpc.  This displacement was obtained for the
model of  ring arms. Its statistical significance is  $2\sigma$.
The displacement increases to 0.5~kpc for the model of trailing
spiral arms even with a small pitch angle, ${i=5^\circ}$. Its
statistical significance  also increases. In contrast, the
displacement decreases for the model of leading arms. The problem
is that  a trailing arm cannot be drawn through the complexes of
young objects in Carina and Cygnus, but  these complexes fall
nicely on a leading arm (Fig.~3).

\section*{Spiral Arms and Spurs}

The main puzzle of the Galactic spiral structure in the solar
neighborhood lies in appreciable nonrandom  deviations from an
ideal spiral pattern. The images of other galaxies  often show
deviations  from a smooth  spiral pattern on short- and
intermediate scales, which are usually called  spurs. The element
of chaos introduced by spurs into the spiral structure of
galaxies can also be used, in principle, to explain the defects
in the Galactic spiral pattern. Let us use the definition of a
spur given by Roberts and Hausman~(1984) as a region of intense
star formation located far from the line of minimum potential and
consider various spur generation mechanisms.

Roberts and Hausman~(1984) showed that the possible long delay of
star formation behind the shock front leads to a blurring of the
 spiral arms identified by a concentration of young stars and
to the appearance of spurs. The successive star formation in the
Galaxy produced  by supernova explosions (Gerola and
Seiden~1978),  also facilitates the formation of spurs. However,
in the presence of a density wave the cloud velocity field  has a
characteristic periodic structure and neither the successive
formation nor the delay of star formation has virtually no effect
on this field (Roberts and Hausman~1984). Moreover, they
facilitate the detection of the  periodic cloud velocity-field
structure by revealing the cloud velocities and distances at
different spiral-wave phases. The fragments of the Carina, Cygnus,
and Perseus arms that we identified  cannot  be  spurs of this
type, by any means. A comparison of their kinematics with the
computed  velocity field (Roberts and Hausman~1984) clearly
indicates that they are located near the line of minimum
potential. The complex of young objects in Sagittarius whose
kinematics or, more specifically, the radial  residual velocity
of young stars  directed away from the Galactic center (Fig.~3)
is suggestive  of its interarm location must be called a spur by
the definition of Roberts and Hausman.

Roberts and Stewart~(1987) found another spur generation
mechanism. They showed that for a large mean free path,
individual fragments of the spiral pattern can temporarily shift
in phase from each other, while the global spiral pattern of the
cloud subsystem  can exhibit ruggedness and gaps. The shift of
the cloud density maximum (collective sloshing) relative to the
potential minimum   is produced by  the tuning of epicyclic
motions of individual clouds (Roberts and Stewart~1987). For
collisionless models the displacement of individual arm fragments
relative to each other can reach $20\%$ of the interarm distance,
while for a system with a mean free path of $p=0.2$~kpc it does
not exceed $5\%$. However, the collective sloshing mechanism can
no longer explain the $0.5-kpc$ displacement of the kinematic
positions of the Carina and Cygnus arm fragments. The Galactic
cloud system  is undoubtedly  collisional and, consequently, this
 mechanism cannot produce large deviations from a smooth spiral
pattern.

Weaver~(1970) found another type of spurs similar to the branches
at the outer arm edges. Elmegreen~(1980) found their mean pitch
angle  to be ${60^\circ\pm10^\circ}$. Weaver~(1970) and
Elmegreen~(1980) believe that the Orion region is such a spur in
our Galaxy. Strong gas compression in the spiral arm, which can
trigger the growth of perturbations in a direction almost
perpendicular to the arm, can be responsible for the appearance
of such spurs (Balbus~1988; Kim and Ostriker~2002). However, the
Carina, Cygnus, and Perseus arm fragments that we identified do
not belong to this type  of spurs either. First, no other spiral
arms  from which these fragments  could appear  as spurs are
observed. Second, the Cygnus and Perseus arm fragments are
definitely elongated in the azimuthal direction (Fig.~3).

Yet another type of spurs in the interarm space was found by
Contopoulos and Grosbell~(1986, 1988) when studying the nonlinear
effects in  high-order resonances between the epicyclic frequency
and the relative rotation rate of the spiral pattern. The spurs
and gaps in the spiral structure near corotation are produced by
nonlinear effects, which lead to difference in the shapes and
orientations of the periodic orbits inside and outside the 4/1
resonance (${\kappa/(\Omega-\Omega_p)=4/1}$). The orbits outside
the 4/1 resonance can be oriented in such a way that the maximum
of their crowding  may not coincide with the spiral arm. However,
nonlinear effects arise in the case of high potential
perturbation amplitudes and must be observed in galaxies with
large arm pitch angles, ${i>30^\circ}$ (Contopoulos and
Grossbell~1986). The tightly would spiral pattern of our Galaxy
(${i=5^\circ}$) cannot, in principle, undergo nonlinear  effects
near corotation.

Thus, none of the spur types listed above can explain the
peculiar features of the Galactic spiral structure, more
specifically, the deviations from an ideal spiral pattern. The
potential perturbation itself is, probably, not  an ideal
monochromatic spiral wave.

Some  authors (Byrd~1983; Byrd et al.~1984) consider the spurs
produced by the gravitation of a massive complex rotating in the
Galactic disk (gravitational spurs). However, such spurs differ
from the spiral-arm fragments only in their length and
perturbation amplitude. Both are produced by a gravitational
potential perturbation  and are located near the potential
minimum.

\section*{Conclusions}

Some of the peculiar features of the periodic velocity-field
structure for  OB associations (Mel'nik et al.~2001) can be
explained by using the model of Roberts and Hausman  (1984), in
which the behavior of a system of dense clouds is considered  in a
perturbed potential. The absence of statistically significant
variations in the azimuthal velocity across the Carina arm,
probably, results from  its sharp increase behind the shock
front, which is easily blurred by distance errors. The existence
of a shock wave in  spiral arms and, at the same time, the
virtually free motion of OB associations in epicycles can be
reconciled in the model of particle clouds with a mean free path
of $0.2\text{--}2$~kpc. The velocity field of OB associations
exhibits two appreciable nonrandom  deviations from an ideal
spiral pattern: a 0.5-kpc displacement  of the Cygnus- and
Carina-arm fragments from one another and a weakening  of the
Perseus arm in  quadrant III. Nevertheless, the identified
fragments of the Carina, Cygnus, and Perseus arms do not belong
to any of the known  types of spurs. The perturbation of the
Galactic potential  is, probably,not a monochromatic spiral wave.

The spiral arms of the Galaxy are often compared with those of
the Andromeda galaxy. Both galaxies have a tightly wound spiral
pattern and similar streaming motions. The mean interarm distance
of $\lambda=3$~kpc observed in the Andromeda galaxy at distances
$R=6\text{--}16$~kpc corresponds to a mean pitch angle of
$i=7^\circ$ (Arp~1964; Braun~1991). The spiral pattern of the
Galaxy in the solar neighborhood has similar parameters,
$\lambda=2$~kpc and ${i=5^\circ}$ (Mel'nik et al.~1999, 2001).
The change in the velocity of neutral and molecular hydrogen
clouds across the spiral arms in the Andromeda galaxy is
10--20~km/s (Braun~1991; Neininger et al.~2001), which is close
to the velocity of streaming motions of young stars in our
Galaxy. In addition, the distribution of HII regions in the
Andromeda galaxy also exhibits deviations of individual arm
fragments from an ideal monochromatic spiral pattern (Considere
and Athanassoula~1982; Braun~1991).

\subsection*{Acknowledgments}

I am grateful to A.V.Zasov for a discussion and helpful remarks,
which contributed to a substantial improvement of the paper. I
wish to thank I.I.Pasha, A.S.Rastorguev, and Yu.N.Efremov for
useful advice and remarks. The work was supported by the Russian
Foundation for Basic Research (projects nos.~02-02-16667 and
00-02-17804), the Council for the Program of Support of Leading
Scientific Schools (grant no.~00-15-96627), and Federal Research
and Technology Program "Astronomy" .

\subsection*{References}

%\begin {enumerate}

 \noindent Adler D.S., and Roberts W.W., Astrophys. J. {\bf 384}, 95 (1992).

 \noindent  Arp H.,  Astrophys. J. {\bf 139}, 1045 (1964).

 \noindent  Balbus S.A., Astrophys. J. {\bf 324}, 60 (1988).

 \noindent  Blaha C., and Humphreys R.M.,  Astron. J. {\bf 98}, 1598 (1989).

 \noindent  Braun R., Astrophys. J. {\bf 372}, 54 (1991).

\noindent  Byrd G.G., Astrophys. J. {\bf 264}, 464 (1983).

 \noindent  Byrd G.G.,  Smith B.F., and Miller R.H., Astrophys. J.
 {\bf 286}, 62 (1984).

 \noindent  Considere S., and Athanassoula E., Astron. Astrophys. {\bf 111},
28 (1982).

 \noindent  Contopoulos G., and Grosbol P.,
 Astron. Astrophys. {\bf 155}, 11 (1986).

 \noindent  Contopoulos G., and Grosbol P., Astron. Astrophys. {\bf 197},
83 (1988).

 \noindent  Dambis A.K., Mel'nik A.M., and Rastorguev A.S.,
 Astron. Letters, {\bf 21}, 291 (1995).

 \noindent  Dambis A.K., Mel'nik A.M., and Rastorguev A.S.,
  Astron. Letters, {\bf 27}, 58 (2001).
  ftp://lnfm1.sai.msu.ru/pub/PEOPLE/rastor/Dambis et al-2001-OB associations (eng).pdf

 \noindent   Elmegreen D.M., Astrophys. J. {\bf 242}, 528 (1980).

 \noindent  Gerola H., and Seiden P.E., Astrophys. J. {\bf 223}, 129 (1978).

 \noindent  Glushkova E.V., Dambis A.K., Mel'nik A.M.,
 and Rastorguev A.S., Astron. Astrophys. {\bf 329}, 514 (1998).

 \noindent  Hausman M.A., and Roberts W.W., Astrophys. J. {\bf 282}, 106
(1984).

 \noindent  Kim W.-T., and Ostriker E.C., Astrophys. J. {\bf 570}, 132
(2002).

 \noindent  Kwan J., and Valdes F., Astrophys. J. {\bf 315},
 92 (1987).

 \noindent  Levinson F.H., and Roberts W.W.,
 Astrophys. J. {\bf 245}, 465 (1981).

 \noindent  Lin C.C., Yuan C., and Shu F.H.,
 Astrophys. J. {\bf 155}, 721 (1969).

 \noindent  Mel'nik A.M., Dambis A.K.,  and Rastorguev A.S.,
 Astron. Letters, {\bf 25}, 518 (1999).

 \noindent  Mel'nik A.M., Dambis A.K., and Rastorguev A.S.,
 Astron. Letters, {\bf 27}, 521 (2001).
 ftp://lnfm1.sai.msu.ru/pub/PEOPLE/rastor/Melnik et al-2001-Periodic pattern of OB-associations.pdf

 \noindent  Neininger N., Nieten Ch., Guelin M. et al., {\it Proceedings
 of the 205th Symp. of the IAU Galaxies and
Their Constituents at the Highest Angular Resolutions}, Ed. by
R.T.Schilizzi ,  S.Vogel  et al., (PASP, San Francisco, 2001), p.
352.

 \noindent  Pringle J.E., Allen R.J., and Lubow S.H., MNRAS {\bf 327}, 663
(2001).

 \noindent  Rastorguev A.S., Pavlovskaya E.D., Durlevich O.V., Filippova
A.A., Astron. Letters, {\bf 20}, 591 (1994).

 \noindent  Roberts W.W., Astrophys. J. {\bf 158}, 123 (1969).

 \noindent  Roberts W.W., and Hausman M.A., Astrophys. J. {\bf 277}, 744
(1984).

 \noindent   Roberts W.W., and Stewart G.R., Astrophys. J. {\bf 314}, 10
(1987).

 \noindent  Solomon P.M., Sanders D.B., and Rivolo A.R.,
 Astrophys. J. Lett. {\bf 292}, L19 (1985).

 \noindent  Weaver H., in Interstellar Gas Dynamics, ed. Habing H.,
 (Dordrecht: Reidel), IAU Symp. {\bf 39}, 22 (1970).

\noindent  Zabolotskikh M.V, Rastorguev A.S.,  Dambis A.K.,
 Astron. Letters, {\bf 28}, 454
 (2002). ftp://ftp.sai.msu.su/pub/groups/cluster/rastor/AL454.pdf

\noindent \\see my Homepage at
http://lnfm1.sai.msu.ru/{$^\sim$}anna

%\end {enumerate}

%---Figure 1.
\begin{figure}[t]
\includegraphics{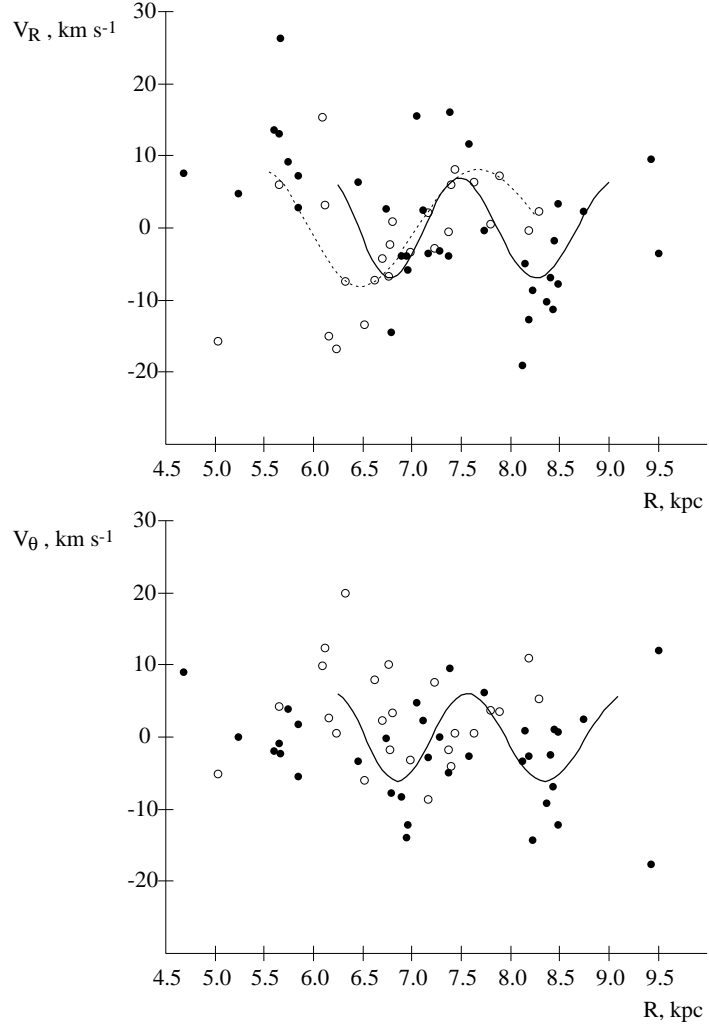} \vspace{14.0cm} \caption{The distribution of the (a)
radial $(V_R)$ and (b) azimuthal $(V_\theta)$  residual
velocities of OB-associations  in Galactocentric distance $R$. OB
associations located in the regions ${0<l<180^\circ}$ and
${180<l<360^\circ}$ are represented by filled and open circles,
respectively. The sine-wave parameters were determined by solving
the equations for line-of-sight velocities and proper-motions in
two regions ${30<l<180^\circ}$ (solid line) and
${180<l<360^\circ}$ (dotted line)  \hfill}
\end{figure}

%---Figure 2.
\begin{figure}[t]
\includegraphics{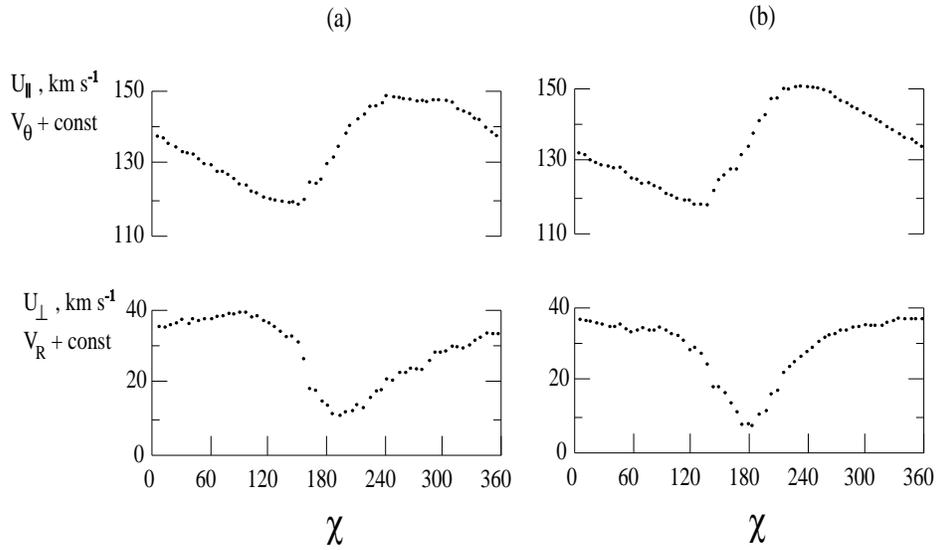} \vspace{7.0cm} \caption{ Variations in the cloud
velocity components perpendicular ($U_\perp$) and parallel
($U_\parallel$) to the shock front with the spiral-wave phase
$\chi$ for cloud systems with a mean free path of (a)~$p=0.2$~kpc
and (b)~$p=1.0$~kpc (Roberts and Hausman~1984, Fig.~13). \hfill}
\end{figure}

%---Figure 3.
\begin{figure}[t]
\includegraphics{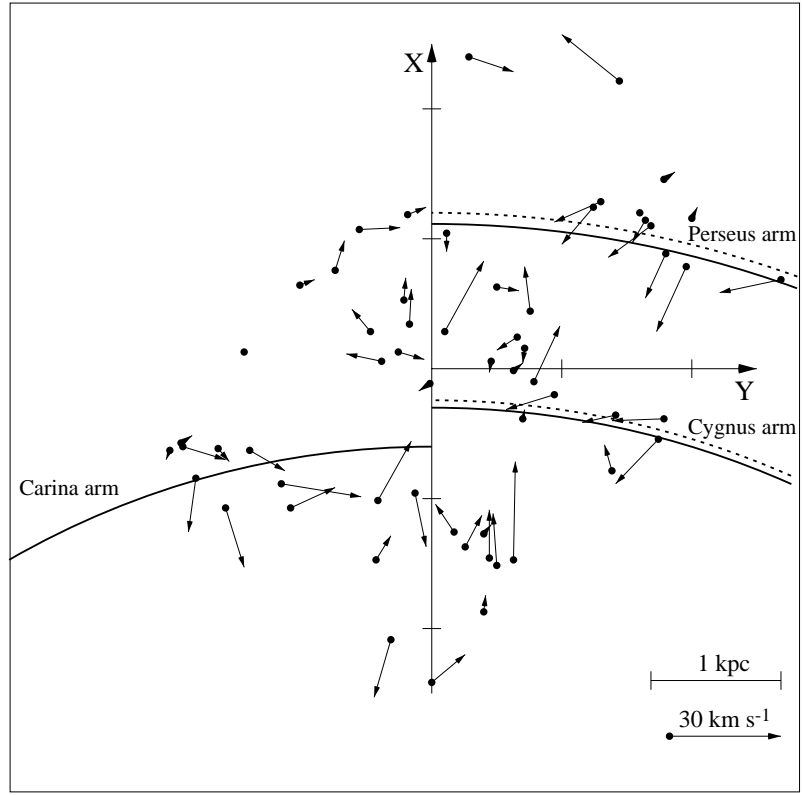} \vspace{8.0cm} \caption{The  residual-velocity field
 for OB-associations in projection onto the Galactic plane. The
$X$ axis is directed away from the Galactic center and $Y$ axis is
in the direction of Galactic rotation. The Sun is at the
coordinate origin. The circular arcs correspond to the minimum
radial (solid line) and azimuthal (dashed line)  residual
velocities of OB associations and determine the kinematic
positions of the Carina, Cygnus, and Perseus arms.\hfill}
\end{figure}

\end {document}